\documentclass[12pt]{iopart}
\usepackage{multirow} 
\usepackage[figuresright]{rotating}
\pdfoutput 1
\begin{document}
\title{Measurement of the two-photon absorption cross-section of liquid Argon with a Time Projection Chamber}

\author{ I.~Badhrees, A.~Ereditato, I.~Kreslo, M.~Messina, U.~Moser, B.~Rossi, M.S.~Weber, M.~Zeller}
\address{Albert Einstein Centre for Fundamental Physics\\Laboratory for High Energy Physics (LHEP)\\
  University of Bern,  Switzerland}
\ead{biagio.rossi@lhep.unibe.ch}
\author{ C.~Altucci, S.~Amoruso, R.~Bruzzese, R.~Velotta}
\address{Dipartimento di Scienze Fisiche\\
Universit\`a di Napoli "Federico II"\\Napoli, Italy}

\begin{abstract}
  This paper reports on laser induced multiphoton ionization at 266 nm of liquid Argon in a Time Projection Chamber (LAr TPC) detector. The electron signal produced by the laser beam is a formidable tool for the calibration and monitoring of the next generation large mass LAr TPCs. The detector we have designed and tested allowed us to measure the two-photon absorption cross-section of LAr with unprecedented accuracy and precision: $\sigma_{ex}=(1.24\pm0.10_{\mathrm{stat}}\pm0.30_{\mathrm{syst}})\cdot10^{-56}~\mathrm{cm^4s}$. 
\end{abstract}
\maketitle
\section{Introduction}

The liquid Argon Time Projection Chamber (LAr TPC) detector technology allows for uniform and high resolution imaging of massive detector volumes. The operating principle of the LAr TPC is based on the undistorted transport of tracks of ionization electrons in highly purified LAr by a uniform electric field over distances up to a few metres. Imaging is provided by wire planes or other read-out devices placed at the end of the drift path. 
The drifting electrons are collected by the outermost wire plane which gives position and calorimetric information. Additional planes with different orientation can be positioned in front of the collection plane to record the signal induced by the passage of the drifting electrons.   
This provides different projective views of the same event thus allowing space point reconstruction. The coordinate parallel to the drift field is given by the measurements of the drift time given by the time interval between the passage of the ionizing particle in the active volume ($t_0$) and the arrival of the drifting electrons on the wire planes. The $t_0$ can come from the detection of the scintillation light of liquid Argon by means of photomultipliers, or from an external source. The momentum of an incoming particle is inferred by the measurement of its multiple scattering \cite{[scattering]}, while the detection of the local energy deposition provides particle identification. For more information, the reader is referred to Ref.~\cite{[MyLAr]} and references therein quoted. 

The purity of the liquid Argon is a key ingredient to achieve imaging over long drift distances. A purity corresponding to less than 0.1~ppb of electronegative elements such as Oxygen, has to be achieved in order to allow for electron lifetimes of milliseconds and, hence, drift distances of meters, as envisioned for large scale applications of this technique \cite{[7],[8],[9],[10],[13]}. This can be obtained by means of commercial purification cartridges and by Argon recirculation.

The purity of LAr has to be monitored continuously while the TPC is running, thus calling for practical and efficient purity monitoring techniques. To this end, laser based methods are very interesting as they provide a reliable route to tackle the problem. In addition, a laser beam is an ideal tool for performing the energy calibration of the detector, for measuring the drift velocity, the space and time resolution of the detector, the capability of double track separation, the diffusion of the drifting electrons, and also for verifing the homogeneity of  the drift field.     

In this paper we report the analysis of UV laser multiphoton ionization of liquid Argon with the $4^{th}$ harmonic of a Nd-YAG laser. We used a LAr TPC both as a target medium and detector of the electrons produced by the laser radiation.

The interaction of laser light with liquid matter is comparatively less studied with respect to gas and solid phase. Nevertheless, some important features are well understood \cite{[Schmidt]}. Laser ionization can occur as a multiphoton process through the simultaneous absorption of two or more photons via virtual states in a medium. This process requires high photons flux, namely, pulsed lasers. In a multiphoton process, bound electrons absorb several laser photons simultaneously, thus exciting the atom to high energy levels or even to ionization. 

The sole measurements of multiphoton ionization in LAr reported so far were carried out by Sun et al. in 1996 \cite{[Sun]}. An accurate measurement of the two-photon absorption cross section of LAr ($\sigma_{ex}$), which is the atomic intermediate step to ionization, is fundamental for reliable calibration of LAr TPCs. In \cite{[Sun]}, an estimate of the range of $\sigma_{ex}$ is given. Improvements of front-end electronics and DAQ systems of the last 15 years and the increase in knowledge of the LAr technology allows today to measure the two-photon absorption cross section of LAr with unprecedented accuracy and precision.  
                                                          
The measurements described in this paper have been performed with the LAr TPC described in \cite{[MyLAr]}. This was part of a R\&D project we are carrying out in Bern in view of the realization of LAr TPC detectors of increasing size. Results have also been published on the study of a TPC built and operated employing a mixture of LAr and Nitrogen \cite{[LN]}. 
%
\section {Experimental setup}
\label{sec:laser}
\label{sec:data_taking2}
\indent

\begin{figure}
\begin{center}
\includegraphics[width=14cm]{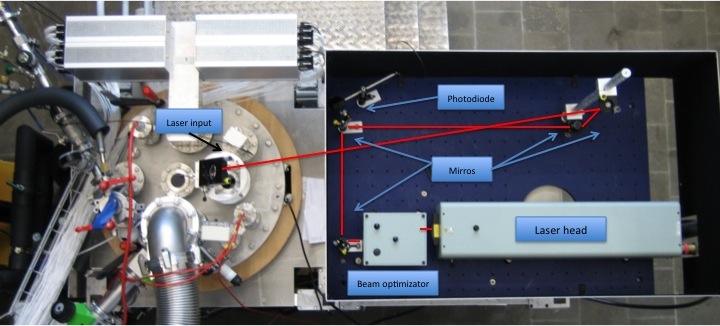}
\end{center}
\caption{\small Photograph of the experimental setup (top view) including the laser beam. Red lines show the beam path before being steered into the detector.} 
\label{fig:laser_setup}
\end{figure}

\begin{figure}
  \center\includegraphics[width=10cm]{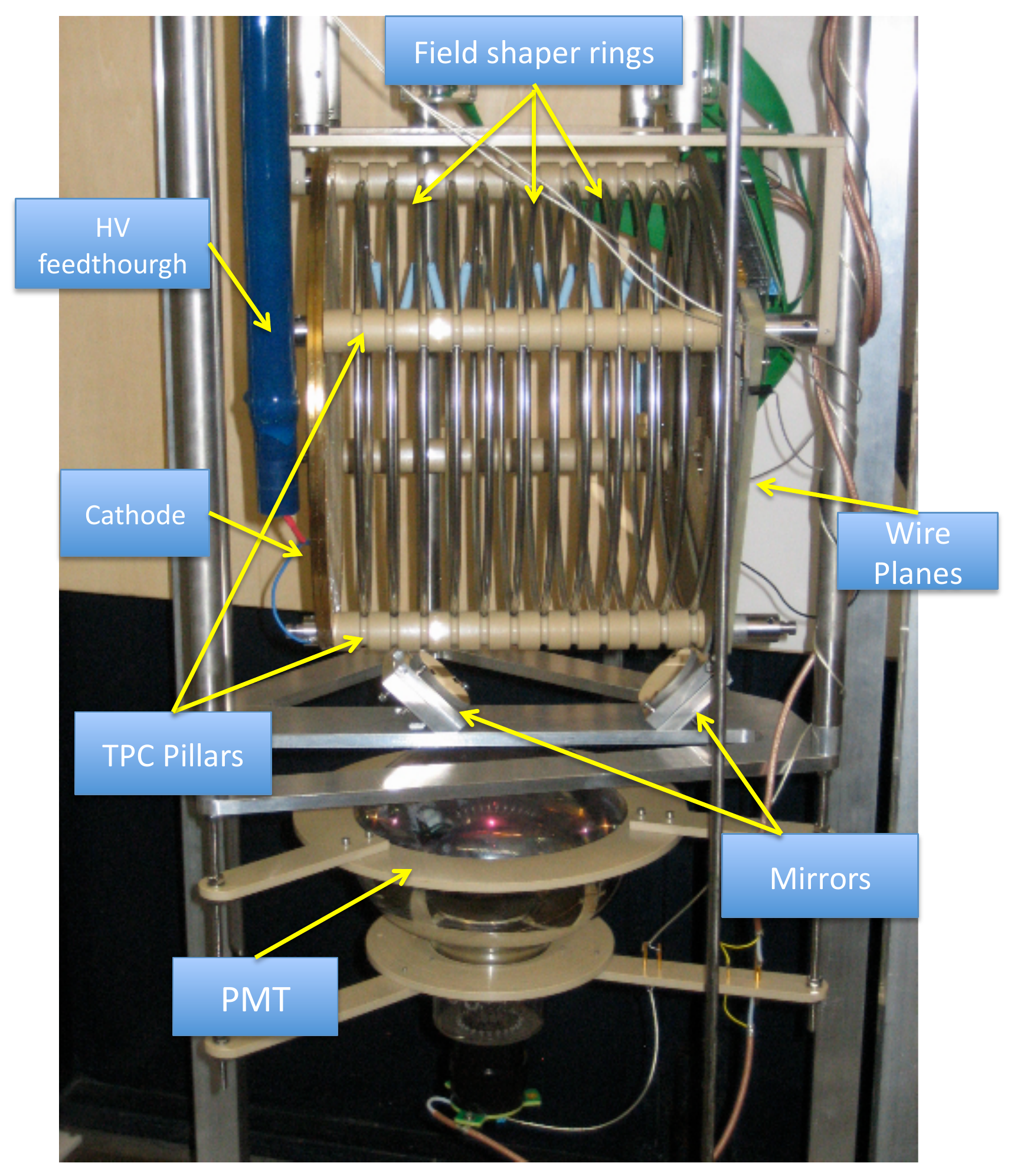}
  \caption{\small Photograph of the inner part of the detector. The TPC, the PMT, the HV feedthrough and two mirrors installed on a stainless steel support are visible. The drift direction is horizontal. }
  \label{fig:TPC_h}
\end{figure}

\begin{figure}
\center\includegraphics[width=12cm]{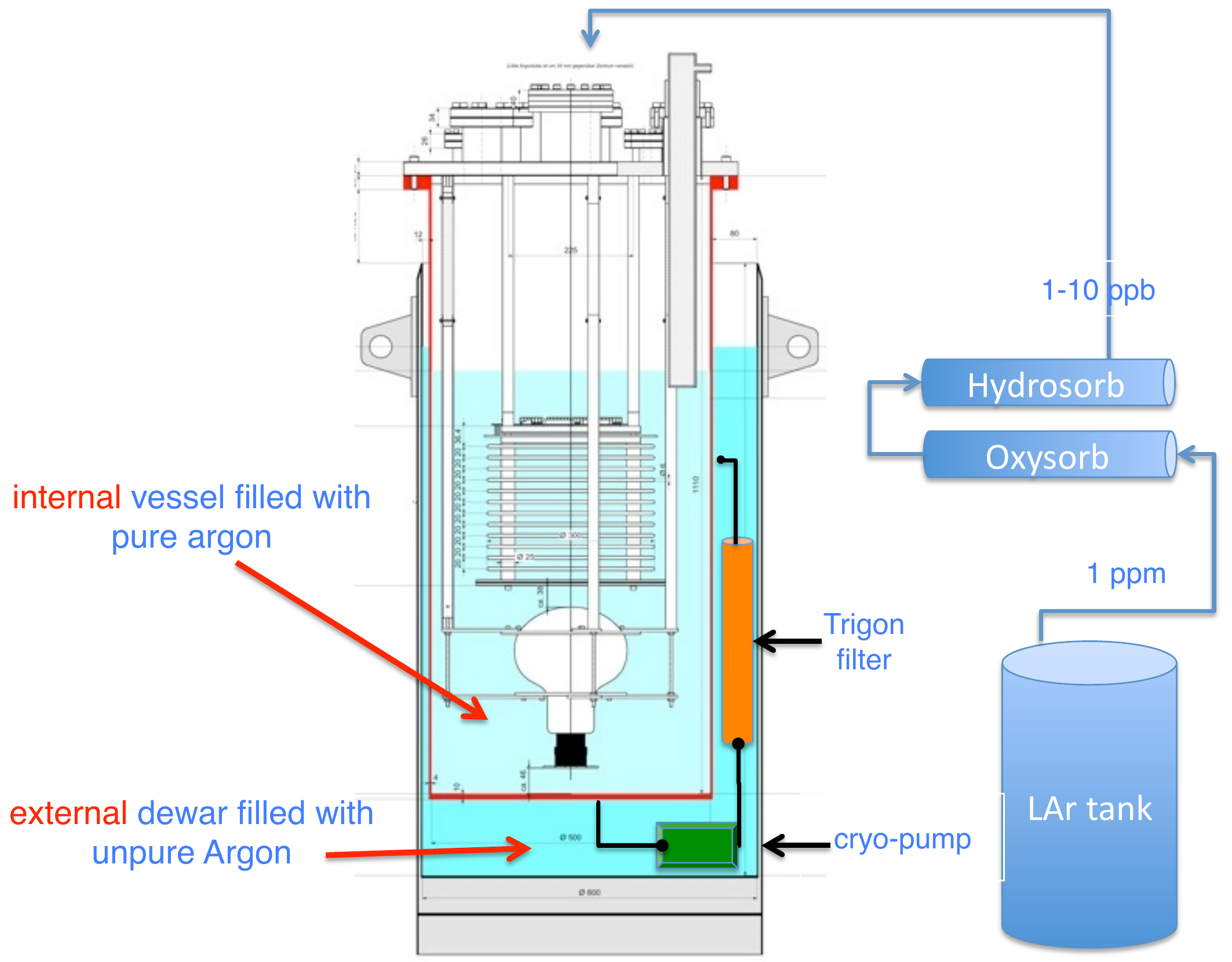}
\caption{Schematic drawing of the detector set-up. The TPC, mounted with the drift coordinate vertically.
The PMT and the relative supports are housed in a cylindrical tube filled with pure Argon. The external vessel that contains un-purified Argon hosts the liquid purification/recirculation system. } 
\label{fig:dewar5}
\end{figure}

The detector (figure~\ref{fig:laser_setup}) used for the measurements consists of a TPC (figure~\ref{fig:TPC_h}) housed in a cylinder filled with purified liquid Argon, in thermal contact with a bath of liquid Argon (figure~\ref{fig:dewar5}). The TPC is complemented by an electronic read-out and data acquisition system (DAQ), a liquid Argon recirculation and purification system, a photomultiplier (PMT), a series of ancillary equipments for monitoring and control, and a UV-laser with the related optics to allow the study of multiphoton ionization of liquid Argon. 

The TPC detector is contained in a stainless-steel cylindrical tube. The latter measures 50~cm in diametre and 110~cm in height, for an inner volume of about 200~litres. The tube is placed in thermal contact with a bath of liquid Argon and is closed by a stainless-steel flange. A series of additional smaller flanges are positioned on the top flange. They include feed-throughs needed to evacuate the inner volume of the vessel, to fill it with liquid Argon, to read-out the detector signals, to supply the high-voltage, to house the pressure monitors, level metre and safety valves, and to provide the transport of the UV laser beams to the inner part of the TPC. The various components are described in detail in \cite{[MyLAr]}.

\begin{figure}
\begin{center}
\includegraphics[width=14cm]{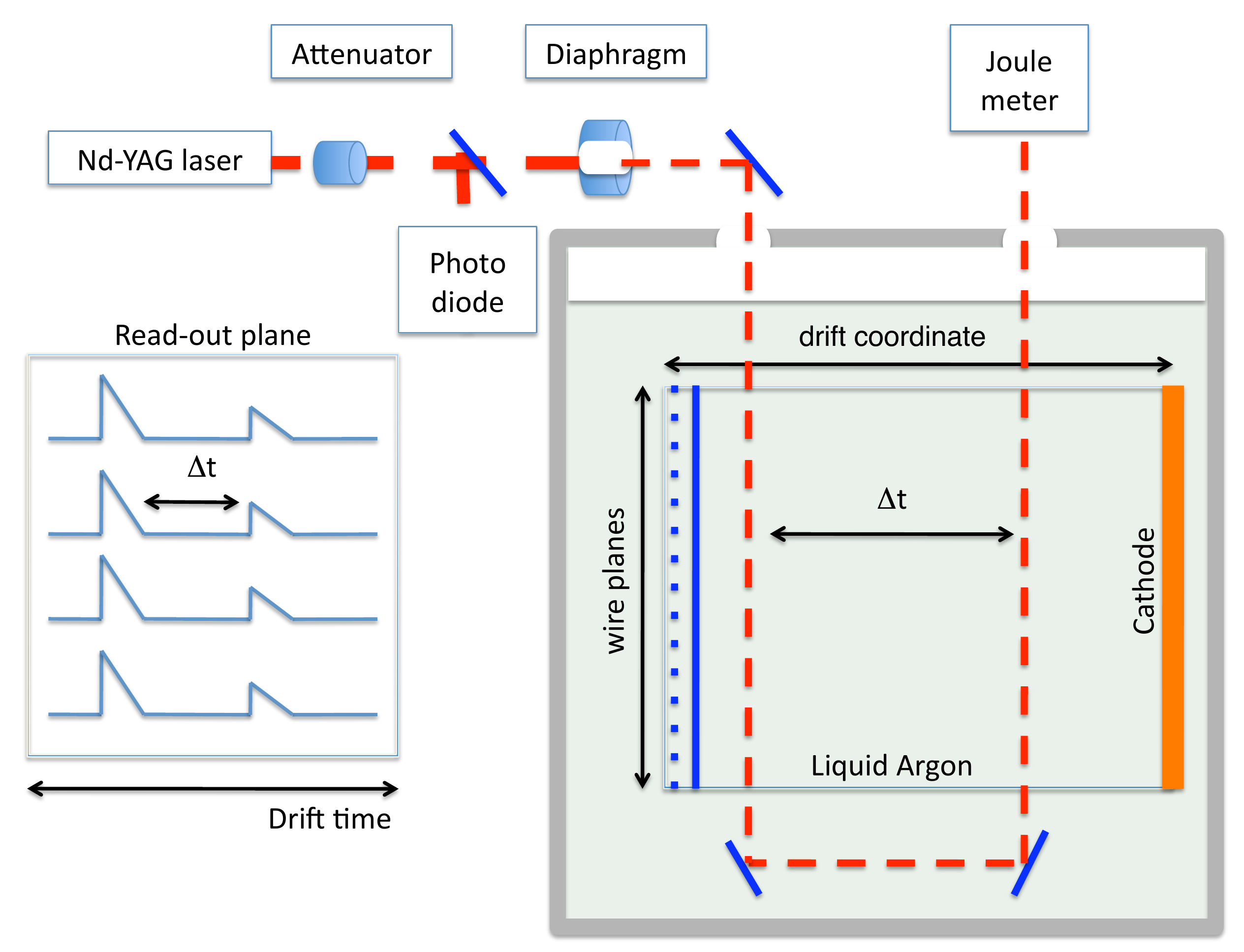}
\end{center}
\caption{\small Schematic view of the laser beam path and the readout response.} 
\label{fig:laser_path}
\end{figure}

Figure~\ref{fig:laser_path} shows the arrangement and the path of the laser beam used for the measurements. 
The laser beam passes through a system of prisms to be {\it cleaned} from the residuals of the $1^{\mathrm{st}}$ and $2^{\mathrm{nd}}$ harmonic components. After a deflection on a 266~nm coated mirror, it is then fed in a variable, motorized attenuator. After the attenuator two parallel beams of variable intensity are present. One is dumped, while the other is used for the measurements. The use of the attenuator is important, as it allows to vary the energy of the laser beam without changing the laser generation parameters. Consequently, the laser shows better stability performance. Both energy and pulse length measurements are performed before the beam is finally steered into the chamber. 

In theoretical calculations (section~\ref{sec:rate_eq}), we assume for simplicity that the transverse distribution of the laser intensity has a square shape. In the experiment, the laser beam passes after the attenuator through a diaphragm of 4~mm diameter selecting the more uniform, central part of the beam. In this way one obtains a definite beam cross section also preventing any effect of the laser beam tails (halo) on the ionization process. In fact, the contribution to ionization of photons in the halo is negligible due to the nonlinear character of the process. 

After the diaphragm, the light is fed into the TPC inner volume by passing through a system of mirrors (figure~\ref{fig:laser_setup}) and a custom made optical feedthrough \cite{[MyLAr]}. 

As described in figure~\ref{fig:laser_path}, the beam passes two times in the active volume, thus producing two parallel tracks at different drift times. After the second passage, the beam is steered out of the dewar by means of another optical feedthrough. The extraction of the laser beam from the TPC allows for the monitoring of its intensity, after ionization of LAr has occurred. 
The attenuation of the intensity due to the optical feedthoughs and mirrors has been separately measured without LAr in the setup.

The measurement of the pulse width is performed by means of a fast photodiode\footnote{Thorlabs DET 10 A.}, calibrated in energy with an energy meter\footnote{GEANTEC - SOLO 2.} and readout by a CAEN V1729 ADC.  The board, a Switched-Capacitor Digitizer, performs the coding of 4 analog channels of bandwidth up to 300~MHz over 12~bits at a sampling frequency of up to 2~GS/s and over a depth of 2520 usable points. It is suited for acquisition of fast analog signals. 
%
\section{Detector operation and data taking}
\label{sec:reco_res}
\indent

Before filling with liquid Argon, the detector vessel is evacuated to a residual pressure of 10$^{-5}$ mbar. The detector is then filled with ultra pure liquid Argon flowed through the purification cartridges in liquid phase, analogous to the purification unit inserted in the recirculation system. The purity of the liquid Argon before passing through the OXYSORB and HYDROSORB \cite{[Messer]} cartridges has a nominal concentration of water and Oxygen of the order of a ppm. The procedure is described in details in Ref.~\cite{[MyLAr]} 

Several measurements were performed with the TPC installed with the electrons drifting along the horizontal direction. The detection of the first UV laser induced ionization tracks have been published in \cite{[MyLAr]}.

\begin{figure}
  \center\includegraphics[width=11cm]{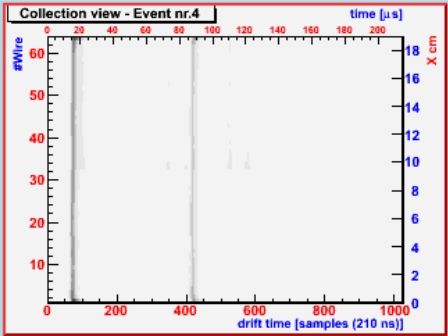}
  \caption{Display of a UV laser ionization event. Two parallel tracks inducing a signal on 64 wires at different drift time (at sample 75 and 420) are visible. The wire number 64 corresponds to the point in which the beam first enters in the TPC.}
  \label{fig:double}
\end{figure}

\begin{figure}
  \center\includegraphics[width=11cm]{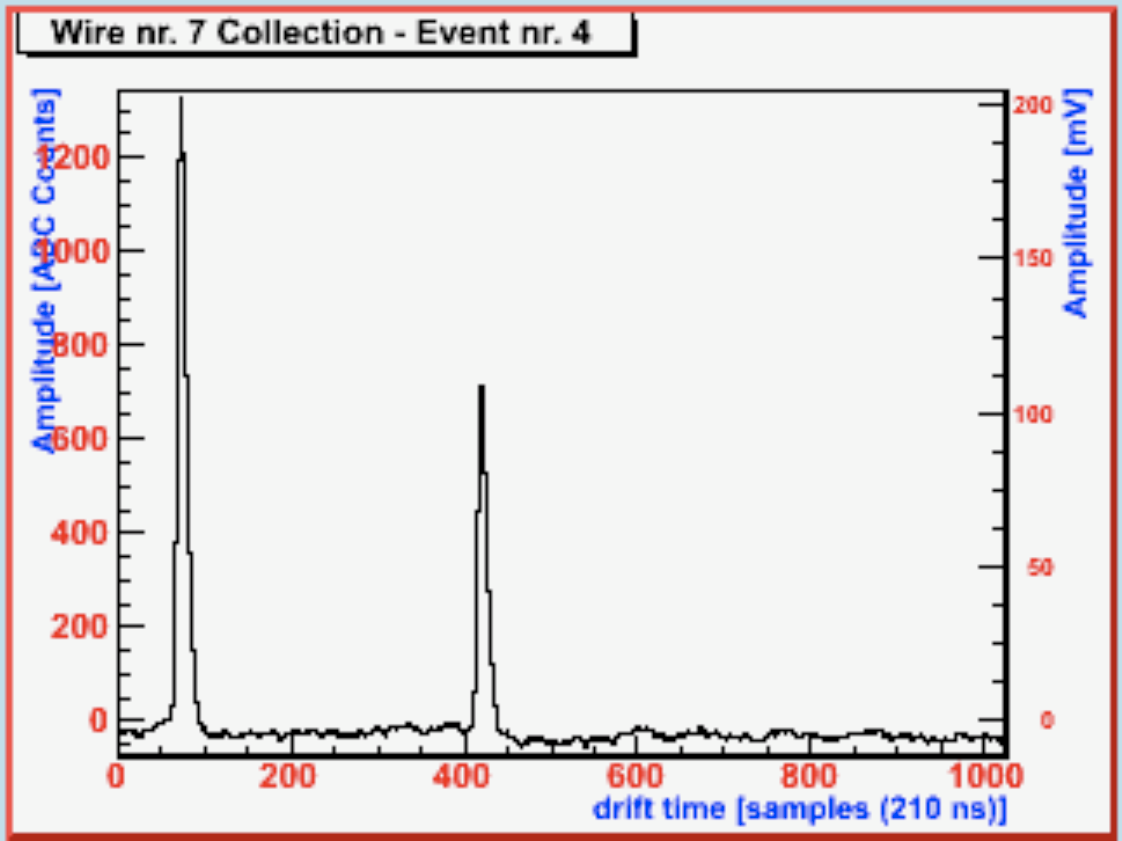}
  \caption{Typical signals induced on a wire of the TPC collection plane by the two UV laser induced tracks. Two peaks at different drift time are clearly visible. The area under each peak provides the electron yield.}
  \label{fig:typ_signal}
\end{figure}

The monitoring of the laser photon flux was performed pulse to pulse. In figure~\ref{fig:double} the display of a typical UV laser induced event is shown. Two parallel tracks along all the wires at different drift time are well visible. The beam first generates the same signal over all the wires of the TPC collection plane at the same time (drift coordinate), and then, after the reflection on two mirrors the beam passes through the active volume again inducing another track parallel to the first one. In figure~\ref{fig:typ_signal} typical waveform peaks induced by a single UV laser pulse on a collection wire are shown. Two peaks at different drift time are clearly visible. In this configuration, for each laser pulse we have 64 independent measurements of the electron yield produced by the first and second track. Combining the information from the two wire planes, a three-dimensional reconstruction of the track is performed. From the collection plane, a calorimetric information about the electron yield by the laser beam is retrieved. 

The distance between the first (second) track and the wire planes is 28~mm (183~mm). From the distance between the laser tracks (155~mm) and the measurement of the difference between the drift time of the two peaks (wire by wire), one can determine the electrons drift speed. Finally, from the ratio of the area under the two peaks we can measure the electron attachment to electronegative impurities and thus, the actual purity of the LAr. 

The electrons attachment to the electronegative impurities present in the LAr during the drift and the charge recombination diminish the charge collected by the wire planes, affecting the reconstruction of the number of electrons produced by laser beams. During the data taking, both LAr purity and charge recombination were measured and, thus, the correction factors were applied to the data presented in section~\ref{sec:results}. 

The concentration of electronegative impurities was at level of 1~ppb of O$_2$ eq., leading to an attenuation length for the drifting electrons of about $\lambda$=650~mm with a drift field of 1~kV/cm, a factor about 20 larger than the distance between the first track and the wire planes.  

\begin{figure}
\begin{center}
\includegraphics[width=10cm]{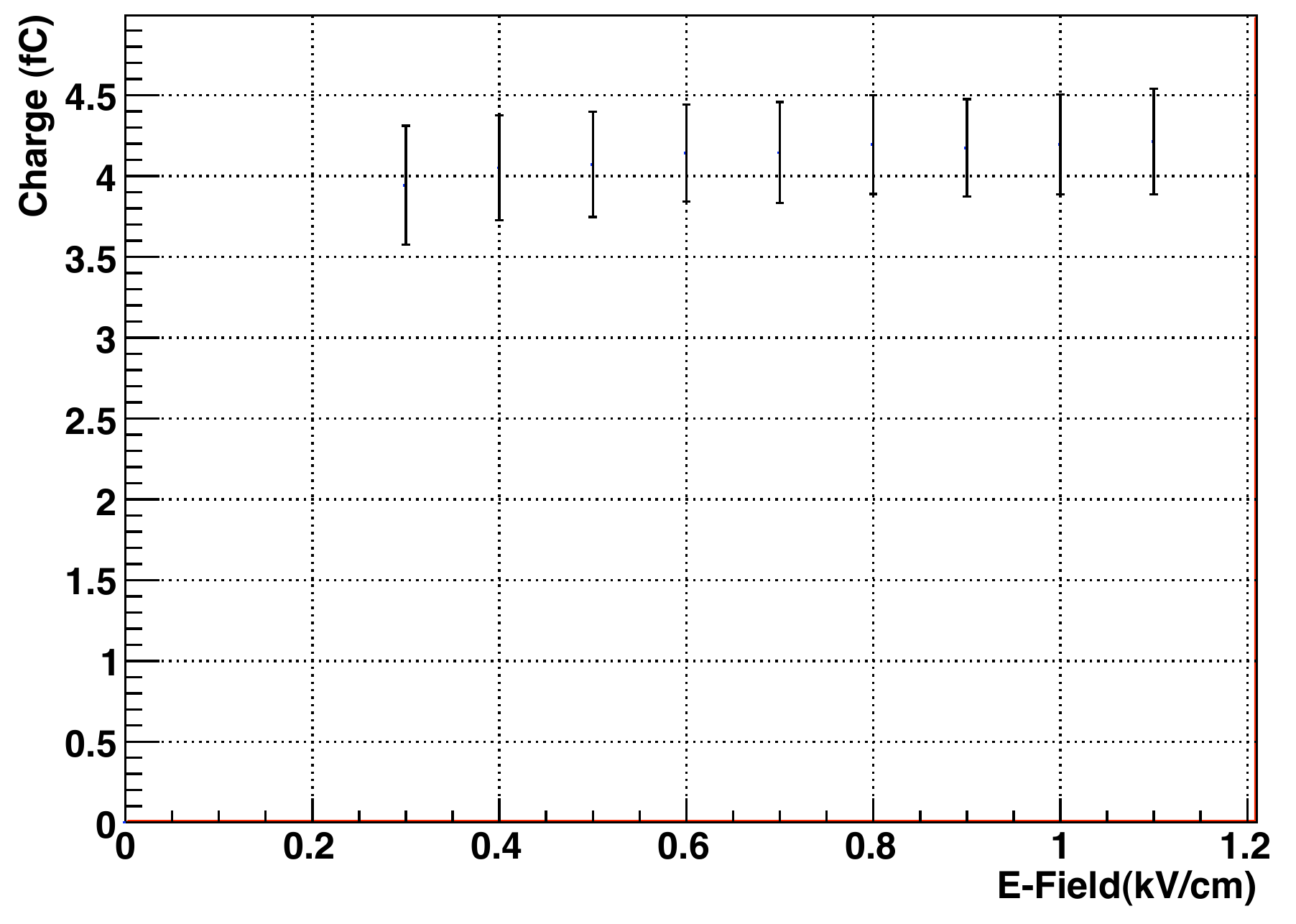}
\end{center}
\caption{\small Plot of the charge produced by the UV laser beam as a function of the drift electric field. This plot shows that the charge recombination is insignificant for the UV laser beam induced events with respect to events with minimum ionizing particles. At the operational drift field value of 1~kV/cm the charge recombination can be neglected.} 
\label{fig:reco}
\end{figure}

The effect of charge recombination was retrieved by the data of Figure 7, which shows the charge collected by a single wire, produced by UV laser beams as a function of the drift electric field. 
For this set of data, the laser repetition rate was 10~Hz and the energy per pulse about 1.0~mJ. Nine different runs of 1000 events for each drift field value were recorded. Data were taken during 10 minutes, in orded to have the same purity conditions for each run. Each entry of the plot of figure~\ref{fig:reco} is the mean value (and the RMS) of the charge distribution for each run. Only information provided by the laser track closest to the wire planes is considered for this analysis. The laser peak-to-peak energy instability (about 5\%), the energy resolution of the TPC, and the uncertainties introduced by the reconstruction algorithm contribute to the ordinate's error bars. 

For drift fields above E$\geq$0.5 kV/cm the data points lie on a flat line indicating that charge recombination for this region is negligible. For drift fields below E$<$0.5 kV/cm an indication of a decrease in the collected charge appears, but no statistically significant statements can be made. 
This shows that the charge recombination is insignificant for the UV laser beam induced events compared to events with ionizing particles for which the recombination at 0.5~kV/cm is about 25\%. 

This difference is likely due to the fact that the charge produced by laser beams is distributed over a much larger volume, i.e. the transverse laser beam spot (diameter $\sim$3~mm) is much larger than the point-like spot of a m.i.p. track. Although the total number of electrons produced by the laser can be large (up to several hundreds of m.i.p. equivalent), the resulting space charge density for laser ionization is, thus, much smaller than for m.i.p., and charge recombination effects may indeed be insignificant.

Since the operational field used for the cross section measurement reported in this paper was 1~kV/cm, we neglect the recombination effect on our data and in the calculation presented in section~\ref{sec:rate_eq}.

The fact that we found negligible charge recombination for laser ionization is interesting for the understanding of the recombination mechanism and is currently matter of further investigations. A detailed study of the possible dependence of the charge recombination from the laser beam energy will be subject of our forthcoming studies.
%
\section{Multiphoton absorption of LAr}
\indent

An atom with ionization energy $\mathrm{E_i}$ can be ionized by irradiation with photons with energy E=h$\nu$ much lower then $\mathrm{E_i}$ if the photon flux is intense enough to promote multiphoton absorption.  

One of the features of multiphoton absorption is that it can occur via laser-induced virtual states, which are not eigenstates of the atom. Thus, it does not require any intermediate atomic state. These laser induced virtual states are characterized by a lifetime $\tau^{virtual}$ of the order of one optical cycle. For the laser used in the present experiment we have:
\begin{equation}
\tau^{\mathrm{virtual}}_{\mathrm{266~nm}} \approx  \frac{\lambda}{2\pi c}\approx 1.4\cdot10^{-16}~\mathrm{s}
\end{equation}
while the lifetime of an intermediate (excited) {\it real} state is of the order of $\tau\sim10^{-8}$~s \cite{[Multibook]}. When an atomic state is located not too far from a laser-induced virtual state (quasi-resonant ionization), the multiphoton absorption process is stronlgy enhanced, thus requiring a laser intensity much lower than in the non-resonant case \cite{[Multibook]}.
%
\subsection{Energy levels in liquid Argon}
\label{sec:LAr_levels}
\indent

\begin{figure}
\center\includegraphics[width=15cm]{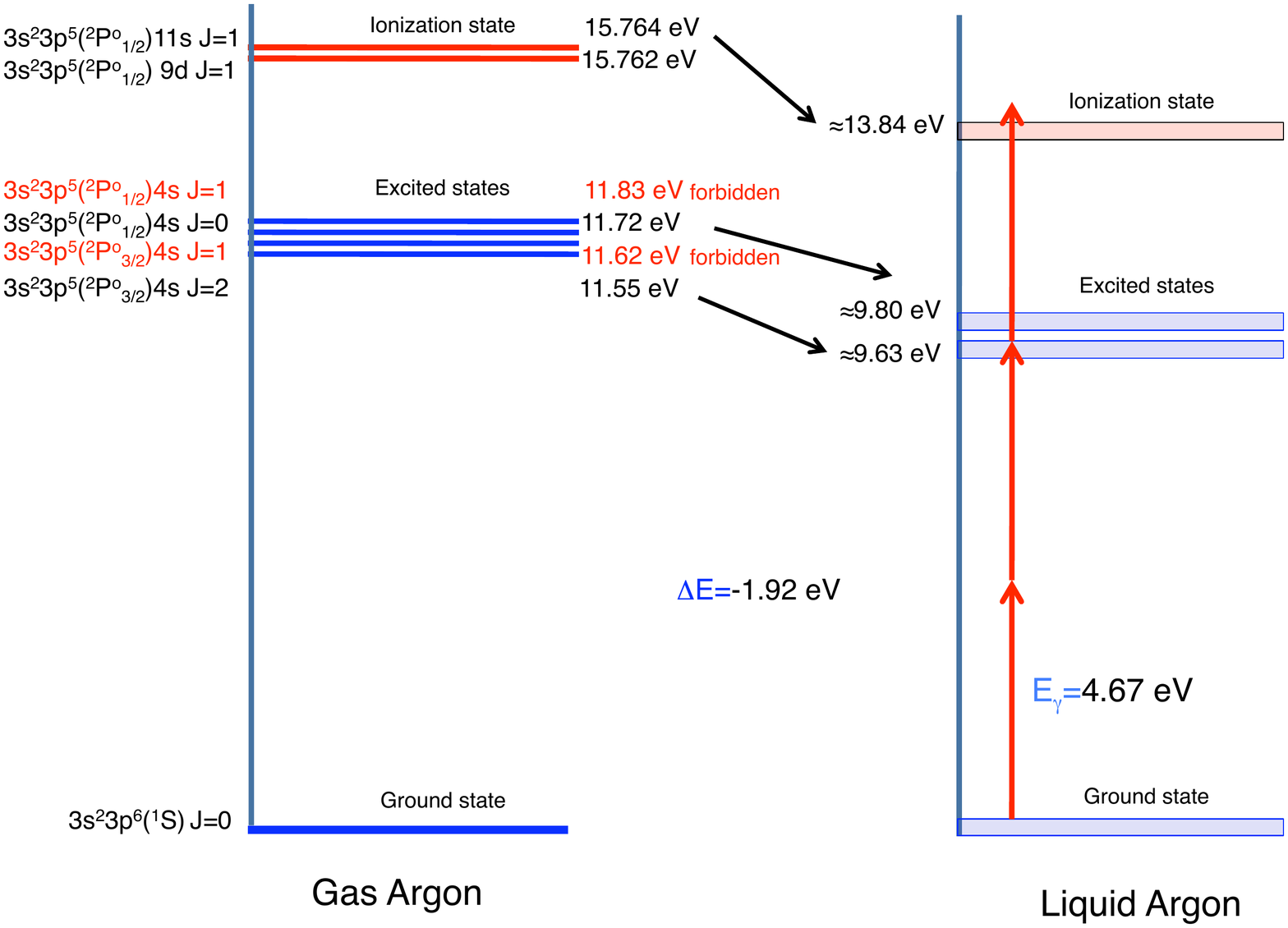}
\caption{\small{Schematics of the gas/liquid Argon energetic levels. In the passage from gas to liquid phase, the absolute values of the energetic levels of the atom are reduced \cite{[Schmidt]} and the lines broaden, becoming energy bands as in the condensed matter.}} 
\label{fig:levels}
\end{figure}

The interaction of laser light with matter in the liquid phase has been comparatively less studied with respect to gas and solid phase. Nevertheless, some important features are well understood \cite{[Schmidt]}. 
In the passage from gas to liquid phase, the energy separation between the energy levels reduces. 
This is sketched in figure~\ref{fig:levels} where the energy levels of gas and liquid Argon are compared. The ionization potential for Argon gas (GAr) is I$_\mathrm{p}$ = 15.76~eV \cite{[NIST]}, and four excited states potentially involved in a multiphoton process lie at 11.83~eV, 11.72~eV, 11.62~eV and 11.55~eV, respectively. For two-photon absorption $\Delta \mathrm{J=0, 2}$ and, hence, only the transitions to levels with J=0 and J=2 are allowed by conservation rules. 
The effects leading to the energy shift can be summarized in the following formula \cite{[shift]}:
\begin{equation} 
I_p^{\mathrm{liquid}}=I_p^{\mathrm{gas}}+P_++V_0
\label{eq:shift}
\end{equation}
where $I_p^{\mathrm{gas}}$ ($I_p^{\mathrm{liquid}}$) is the ionization potential in gas (liquid) phase and $V_0$ is the energy of the quasi-free electron at the bottom of the conduction band of the liquid. The polarization energy $P_+$ of the positive ion is given by the Born equation \cite{[born]}:
\begin{equation} 
P_+=-\frac{e^2}{2r_+}(1-\epsilon^{-1})
\label{eq:born}
\end{equation}
In the classical approximation, a liquid is treated as a uniform continuum with the optical dielectric constant $\epsilon$, while a positive ion is regarded as a sphere with radius $r_+$ and charge $e$. For Argon the values experimentally measured are $\epsilon$=1.5 at 550~nm and $\epsilon$=1.9 at 128~nm, but an extrapolation from a plot published in \cite{[ICA_epsilon]} leads to $\epsilon$=1.59 at 266~nm. Since $V_0=-0.17$~eV \cite{[V0]}, $r_+=0.088$~nm, we have $P_+=-1.75$~eV thereby the ionization state in liquid turns out to be $I_p^{\mathrm{liquid}}=I_p^{\mathrm{gas}}-1.92$~eV. Consequently, the ionization and the lowest excited state energies become, respectively: 
\begin{equation} 
I_p^{\mathrm{liquid}}=13.84~\mathrm{eV} ~~\mathrm{,} ~~P_e^{\mathrm{liquid}}=9.63~\mathrm{eV}
\label{eq:shift_res}
\end{equation}

In our experiment we have used the $4^\mathrm{{th}}$ harmonic ($\lambda$=266~nm, corresponding to a photon energy of 4.67~eV) of a Nd-YAG laser source, the main features of which are reported in Table~\ref{tab:laser}. 
\begin{table}[ht]
\caption{$4^{\mathrm{th}}$ harmonic UV laser specifications (Continuum, model Surelite I-10)}
\centering
\begin{tabular}{  c c c c }
\hline\hline
wavelenght (nm) & max rep. rate (Hz) &  max energy (mJ)  & energy stability  \\ 
         266                &            10                             &         82                    & 5\%\\ 
pulse width (ns)   & rod diameter (mm)           & divergence (mrad) &FWHM (ns)\\
      5-7                    &               6                              &  0.6                          &5.7\\

\hline
\end{tabular}	
\label{tab:laser}
\end{table}

At this wavelength at least three photons are needed to produce ionization in LAr, although the intermediate real state can make the actual multiphoton transition a 2+1 transition. The lowest allowed excited state is estimated to be at 9.63~eV, which is 0.3~eV higher than the energy of two photons at 266~nm. However, one has to consider that atomic energy levels broaden in liquid phase~\cite{[Schmidt]} and that the energy shift of $\Delta E$=-1.92~eV could be underestimated due to approximations underlying our calculation, {\it i.e.} the true shift should be larger (2.2~eV) such to move down the two excited level bands below 9.5~eV, thereby matching the resonant two-photon absorption. It is worth to highlight that such a shift produces an energy gap free to the VUV photon propagation which is in total agreement with the well known transparency of LAr to its own scintillation light at 128~nm ({\it i.e.} 9.7~eV), the latter being  produced by the decay of excimers formed by particle energy loss.
%
\subsection{Rate equations}
\label{sec:rate_eq}
\indent

In a multiphoton ionization process the N-photon ionization rate $W$ is given by $W=\sigma_NF^N$, where $\sigma_N$ is the generalized N-photon ionization cross-section. The rate $W$ is expressed in reciprocal seconds, $\sigma_N$ is expressed in cm$^{2N}$ s$^{N-1}$ and the laser flux $F$ is in photon cm$^{-2}$ s$^{-1}$. 

For gases the one-photon absorption cross-section $\sigma_1$ is typically of the order of $10^{-17}$cm$^2$\cite{[Hilke]}, whereas the two-photon and the three-photon cross-sections are of the order of $\sigma_2=W/F^2\sim10^{-50}\mathrm{cm}^4\mathrm{s}$ and $\sigma_3=W/F^3\sim10^{-83}\mathrm{cm}^6\mathrm{s}^2$, respectively. 

From the energy levels depicted in figure~\ref{fig:levels} we can write the following rate equation to retrieve the expected rate of excited electrons produced:
\begin{equation}
\frac{dN_{ex}}{dt}=(N_0-N_{ex})F^2\sigma_{ex}-N_{ex}\sigma_iF-N_{ex}/\tau
\label{eq:e}
\end{equation}
where $N_0$ is the initial density of atoms in the ground state, $N_{ex}$ the density of atoms in the excited state, $\sigma_{ex}$ the two-photon transition cross-section from ground to the intermediate excited state, $\sigma_i$ is the single photon transition cross-section from the excited state to the continuum, and $\tau$ the lifetime of the excited state, respectively. The rate of ionization is given by:
\begin{equation}
\frac{dN_i}{dt}=N_{ex} \sigma_i F
\label{eq:i}
\end{equation}
where $N_i$ is the density of ionized atoms (or number of extracted electrons). Here we consider the ion/electron recombination (analyzed in section~\ref{sec:reco_res}) negligible for UV laser induced events at our operational drift field. Defining $h=\sigma_{ex}F^2+\sigma_iF+\frac{1}{\tau}$, replacing equation~(\ref{eq:e}) in equation~(\ref{eq:i}) and time-integrating ($\Delta t=s$) one obtains:
\begin{equation}
N_i=\frac{N_0\sigma_i\sigma_{ex}F^3}{h}\Big{(}\frac{e^{-hs}-1}{h}+s\Big{)}
\label{eq:slope}
\end{equation}
where $s$ is the FWHM of the laser pulse (in our case s=($5.7\pm5\%)\cdot10^{-9}~\mathrm{s}$). 

Three different regions can be defined: 
\begin{itemize}
\item low photon flux for $hs<<1$;
\item high photon flux for $hs>>1$, $\sigma_{ex}F^2<<\sigma_iF$ and $1/\tau<<\sigma_iF$;
\item very high photon flux for $hs>>1$, $\sigma_{ex}F^2>>\sigma_iF$ and $1/\tau<<\sigma_{ex}F^2$. 
\end{itemize}
In the region of low photon flux equation~(\ref{eq:slope}) can be written as:
\begin{equation}
N_i=\frac{1}{2}N_0\sigma_{ex}\sigma_is^2F^3
\label{eq:lowslope}
\end{equation}
showing a cubic dependence of the density of the electrons in the ionized state with respect to the flux of photons. In the high photon flux regime equation~(\ref{eq:slope}) becomes:
\begin{equation}
N_i=N_0\sigma_{ex}sF^2
\label{eq:highslope}
\end{equation}
exibiting a quadratic dependence.
Finally, for very high photon flux equation~(\ref{eq:slope}) can be approximated as:
\begin{equation}
N_i=N_0\sigma_i s F
\label{eq:vhighslope}
\end{equation}
with a linear dependence due to the saturation of the two-photon transition.
%
\section{Results and data analysis}
\label{sec:results}
\label{sec:slope_res}
\noindent

In order to estimate the cross-sections $\sigma_{ex}$ and $\sigma_i$ of the rate equations presented above, and the lifetime of the excited state of LAr, we measured the rate of electrons produced by the laser beam as a function of the photon flux. For each laser track we determined the total electron yield produced by the laser (for each wire) simultaneously, monitoring pulse-by-pulse the energy, and FWHM of the beam. For the data set shown below, the beam energy was varied in the range 0.1$\div$2.0~mJ. 
\begin{figure}
\begin{center}
\includegraphics[width=14cm]{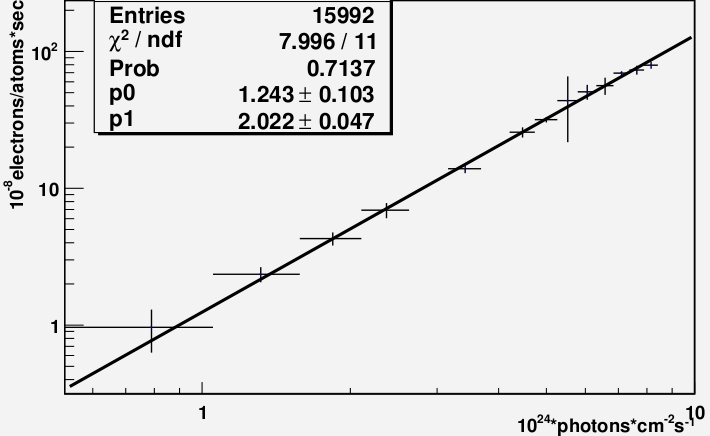}
\end{center}
\caption{\small Profile histogram of the rate of electrons as a function of the photon flux of the laser beam. } 
\label{fig:slope}
\end{figure}

Figure~\ref{fig:slope} reports the rate of electrons (measured in electron atom$^{-1}$ s$^{-1}$) as a function of the photon flux (measured in photon cm$^{-2}$ s$^{-1}$), in a log-log plot. In the investigated intensity range the data are well described by a straight line, and the experimental points are very well fitted with a power law dependence on the photon flux F (see ~\ref{eq:lowslope}-\ref{eq:vhighslope}). In particular, the best fit is obtained by using equation~\ref{eq:highslope} in the following form: 
\begin{equation}
\frac{N_{i}}{N_0s}= \sigma_{ex} \cdot F^n
\end{equation}
with $n$ and $\sigma_{ex}$ as free parameters. The corresponding best fitted values are: 
\begin{equation}
n=2.02\pm0.05
\end{equation}
 \begin{equation}
 \sigma_{ex}=(1.24\pm0.10_{\mathrm{stat}}\pm0.30_{\mathrm{syst}})\cdot10^{-56}~\mathrm{cm^4s}
 \end{equation}
 
The value of $n$ is consistent with 2, indicating that at our laser intensities the limiting step of the three-photon ionization process is the two-photon absorption to the intermediate state. 

The value of $\sigma_{ex}$ presented here provides a reliable and accurate measurements of the two-photon absorption cross section of LAr. It is within the range of the two-photon absorption cross-section for liquids \cite{[Cross_liquids1],[Cross_liquids2]}. The assumption made ($hs>>1$, $\frac{1}{\tau}<<\sigma_iF$ and $\sigma_iF>>\sigma_{ex}F^2$) to approximate equation~(\ref{eq:slope}) to equation~(\ref{eq:highslope}) are justified by the experimental data. Moreover, the limits obtained for $\sigma_i>5\cdot10^{-16}$~cm$^2$ and $\tau>10^{-9}$~s are consistent with the values $\sigma_i=10^{-16}$-$10^{-17}$~cm$^2$ and $\tau=10^{-7}$-$10^{-8}$~s provided by \cite{[Hilke],[Bourotte],[Hurst]}. 

Systematic uncertainties due to the absolute energy calibration of the TPC and to the measurement of the photon flux have been evaluated. We assume a 25\% uncertainty on the absolute number of electrons collected, and 10\% uncertainty on the pulse-by-pulse measurement of the beam energy.   

While our results for $\sigma_i$ and $\tau$ are in reasonable agreement with those reported in \cite{[Sun]}, a disagreement of nearly 3 orders of magnitude is found for the value of $\sigma_{ex}$.
The reason for such a large difference is not completely understood at this stage. However, one can make some considerations. In particular, no uncertainty on the estimation of $\sigma_{ex}$ is given in \cite{[Sun]}, thus the significance cannot be evaluated. As decribed before, in our experimental setup, particular care was taken in the photon flux measurement. The energy of the beam and the FWHM was performed pulse-to-pulse. The energy was measured by two different devices: a fast photodiode placed before the beam was actually steered in the TPC and an energy meter placed at the end of the laser path. The FWHM was monitored by a fast ADC board with 0.5~ns sampling. In this way, we kept the uncertainties low. Moreover, thanks to the spatial reconstruction capabilities of the TPC, we could select for the analysis laser pulses with uniform charge distribution along the track. This is very important for this measurement, since multiphoton absorption is not a linear process. In particular, we also noticed that prior to the installation of the optical feedthrough we observed a large non uniform ionization in the tracks. This is probably due to a focusing/defocusing effect of the laser beam in the passage between the gas/liquid Argon interface inside the detector. Data taken in these conditions provided in average less charge collected per track. 
%
\section {Conclusions and outlook}

We reported on the measurement of UV laser-induced multiphoton ionization of LAr performed with a liquid Argon Time Projection Chamber detector. For the first time, the value of the two-photon absorption cross-section $\sigma_{ex}=(1.24\pm0.10_{\mathrm{stat}}\pm0.30_{\mathrm{syst}})\cdot10^{-56}~\mathrm{cm^4s}$ has been determined with a good accuracy appropriate for quantitative calibration of LAr TPC detectors. The systematic uncertainties on the value of $\sigma_{ex}$ could be reduced even further by improving the accuracy of the TPC calibration.
Further measurements in the range of low photon flux ($hs<<1$) will be performed for a precise determination of the single photon transition cross-section from the excited state to the continuum $\sigma_i$ and the lifetime  $\tau$ of the lower excited state of the LAr. The sensitivity to a lower flux of photons can be improved in several ways. First, with a different position (angle) of the TPC collection wire plane with respect to the laser tracks, in order to increase the portion of charge collected per wire. Second, by improving the present charge preamplifiers ({\it e.g. working in cold}) providing a better signal-to-noise ratio. Third, by using a different readout device with electron amplification, such as the Large Electron Multiplyer (LEM)\cite{[LEM]}. 

The results obtained and described in this paper make us confident on the practical use of UV laser beams technique also for the calibration of LAr TPC detectors, as requested for the envisioned large scale applications of the technique. 
%
%
\section* {Acknowledgements}
The work presented in this paper was conducted thanks to grants from the Swiss National Science Foundation (SNF) and from the University of Bern. We warmly acknowledge both institutions. 
We wish to warmly thank our technical collaborators R.~Haenni, P.~Lutz and F.~Nydegger, for their skillful help in building and operating the detector and the related infrastructure.

\end{document}